\documentclass[conference]{IEEEtran}
\IEEEoverridecommandlockouts
\usepackage{hyperref}
\usepackage[utf8]{inputenc}
\usepackage[T1]{fontenc}
\usepackage{color,soul}
\usepackage{times}
\usepackage{xcolor}
\usepackage{epsfig}
\usepackage{graphicx}
\usepackage{amssymb}
\usepackage{multicol,multirow}
\usepackage{booktabs}
\usepackage{bigstrut}
\usepackage{floatrow}
\usepackage{rotating}
\usepackage{adjustbox}
\usepackage{amsmath}
\usepackage{url}
\usepackage{subfigure}
\usepackage{longtable}
\usepackage{array}

\newcolumntype{L}{>{\centering\arraybackslash}m{3cm}}

\begin{document}

\title{LMBiS-Net: A Lightweight Multipath Bidirectional Skip Connection based CNN for Retinal Blood Vessel Segmentation}


\author{Mufassir M. Abbasi, Shahzaib Iqbal, Asim Naveed, Tariq M. Khan, Syed S. Naqvi, Wajeeha Khalid 

\thanks{Mufassir M. Abbasi and Shahzaib Iqbal are with the Department of Electrical Engineering, Abasyn University Islamabad Campus (AUIC), Islamabad, Pakistan (e-mail: \{muffasir.matloob, shahzeb.iqbal,\}@abasynisb.edu.pk)}  

\thanks{Asim Naveed with the Department of Computer Science and Engineering, University of Engineering and Technology (UET) Lahore, Narowal Campus, Pakistan (e-mail: asim.naveed80@gmail.com)}  

\thanks{Tariq M. Khan is with the School of Computer Science \& Engineering, UNSW, Sydney, Australia (e-mail: tariq.khan@unsw.edu.au)}  

\thanks{Syed S. Nqvi is with the Department of Electrical and Computer Engineering, COMSATS University Islamabad (CUI), Islamabad, Pakistan (e-mail: saud\_naqvi@comsats.edu.pk)}

\thanks{Wajeeha Khalid is with the Department of Pharmaceutics, Shifa Tameer-e-Millat University, Islamabad, Pakistan (e-mail: wajeeha.scps@stmu.edu.pk)}
}

\maketitle

\begin{abstract}
Blinding eye diseases are often correlated with altered retinal morphology, which can be clinically identified by segmenting retinal structures in fundus images. However, current methodologies often fall short in accurately segmenting delicate vessels. Although deep learning has shown promise in medical image segmentation, its reliance on repeated convolution and pooling operations can hinder the representation of edge information, ultimately limiting overall segmentation accuracy. In this paper, we propose a lightweight pixel-level CNN named LMBiS-Net for the segmentation of retinal vessels with an exceptionally low number of learnable parameters \textbf{(only 0.172 M)}. The network used multipath feature extraction blocks and incorporates bidirectional skip connections for the information flow between the encoder and decoder. Additionally, we have optimized the efficiency of the model by carefully selecting the number of filters to avoid filter overlap. This optimization significantly reduces training time and enhances computational efficiency. To assess the robustness and generalizability of LMBiS-Net, we performed comprehensive evaluations on various aspects of retinal images. Specifically, the model was subjected to rigorous tests to accurately segment retinal vessels, which play a vital role in ophthalmological diagnosis and treatment. By focusing on the retinal blood vessels, we were able to thoroughly analyze the performance and effectiveness of the LMBiS-Net model. The results of our tests demonstrate that LMBiS-Net is not only robust and generalizable but also capable of maintaining high levels of segmentation accuracy. These characteristics highlight the potential of LMBiS-Net as an efficient tool for high-speed and accurate segmentation of retinal images in various clinical applications.
\end{abstract}

\vspace{0.5\baselineskip}



\section{Introduction}
\label{intro}

\IEEEPARstart{R}{etinal} images are used to capture the structure of retinal blood vessels, and analyzing their morphology is essential to diagnose pathological conditions \cite{khan2021residual,khan2021rc}. The study of retinal pathology offers valuable information on various ocular diseases \cite{khan2020shallow,khan2020semantically}. Pathological conditions such as diabetic retinopathy and hypertension have the potential to induce substantial modifications in the intricate configuration of retinal vessels \cite{mohite2023retinal,khan2020exploiting}, ultimately causing variations in vessel width and tortuosity. Examination of the distribution of the density of the vessel provides significant information on various underlying medical conditions, due to the interlinked relationship between retinal pathology and the dynamic evolution of the morphology of the retinal vessel \cite{naveed2021towards}. Consequently, the segmentation of retinal blood vessels assumes a critical role in the timely detection of associated diseases.

Segmentation of retinal blood vessels is a challenging task due to their intricate and asymmetric structures of various shapes \cite{imtiaz2021screening}. Current retinal vessel segmentation methods can be broadly categorized into two types: manual segmentation and algorithmic segmentation \cite{naqvi2019automatic}. Manual segmentation involves ophthalmologists manually marking dynamic vessels, but it is a laborious, time-consuming, and costly process due to the complexity of the retinal vessels. The accuracy of manual retinal blood vessel segmentation is highly dependent on the expertise of the clinician. However, this approach is subjective, leading to variations in results between different clinicians and affecting the precision and precision of the segmentation \cite{khan2022t}. Therefore, an imperative requirement for an automated and precise retinal vessel segmentation framework is emerging in the context of computer-aided diagnosis (CAD), with the aim of effectively addressing the aforementioned challenges.

\begin{figure*}
    \centering
    \includegraphics[width=0.9\textwidth]{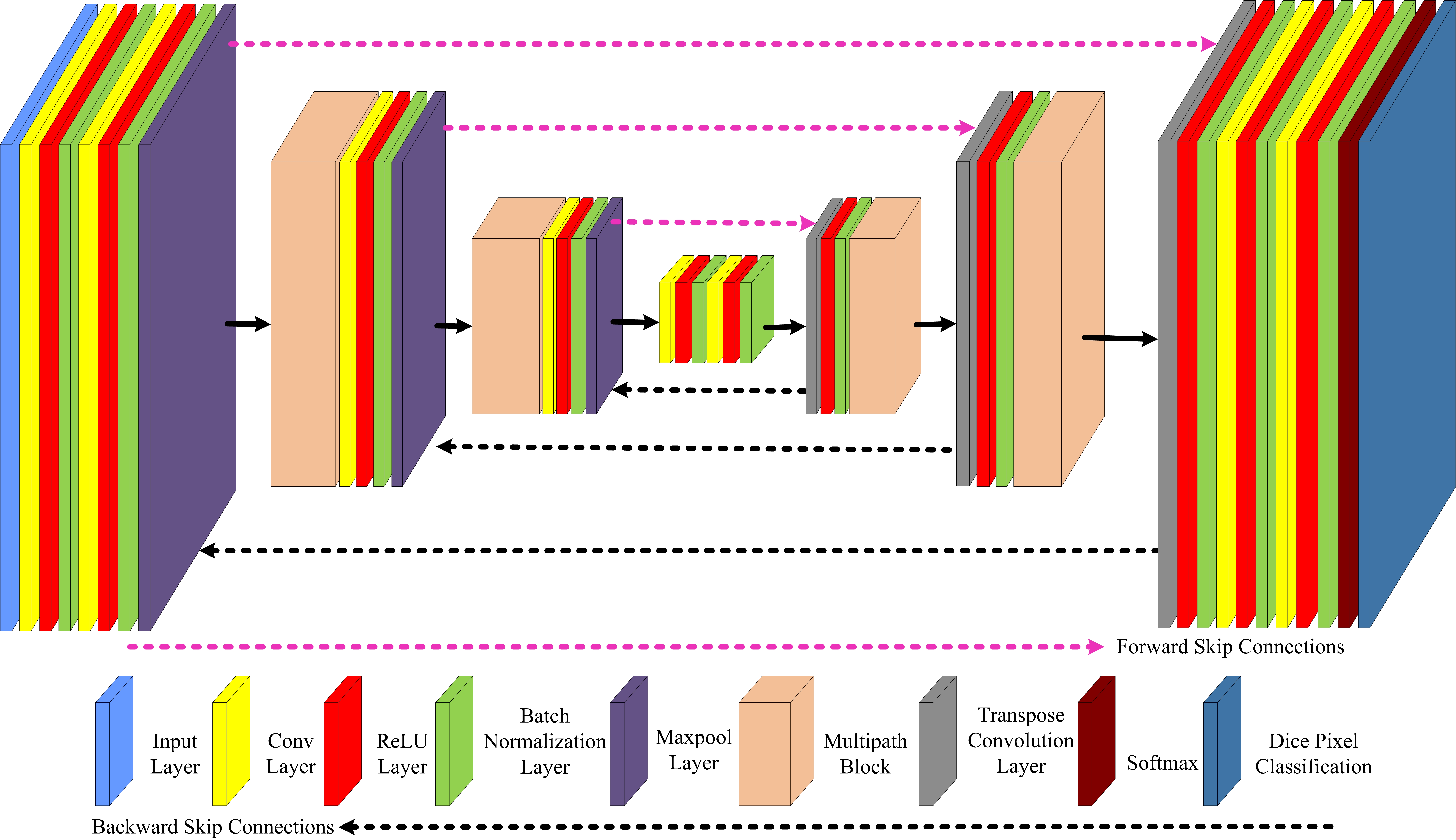}
    \caption{Block diagram of the proposed LMBiS-Net.}
    \label{fig:Model}
\end{figure*}

To develop an automated segmentation framework for retinal vessels, it is crucial to address specific limiting factors. The quality of fundus images plays an important role in extracting feature representations \cite{iqbal2023robust,khalid2023advancing}. Low-contrast fundus images pose challenges, as they hinder the extraction of structural features from the retina \cite{khan2019boosting}. Effective capture of spatial information becomes vital due to the complex and dynamic morphology of retinal vessels, especially when dealing with vessels of varying thickness \cite{khan2020region}. Accurate extraction of slender vessels against intricate fundus image backgrounds requires the augmentation of high-level features. The quality of the resulting segmented mask is significantly influenced by these aspects. To establish a reliable and precise automated retinal vessel segmentation framework, meticulous attention must be devoted to mitigating these constraining factors \cite{iqbal2023mlr}.

In recent years, there has been a surge in deep segmentation models based on U-Net \cite{iqbal2022recent}, aimed at improving segmentation precision and efficiency in medical image segmentation tasks. These models employ various optimization schemes, such as attention mechanisms \cite{yang2022attention}, feature aggregation \cite{khan2023feature, khan2023retinal}, recurrent convolution \cite{alom2019recurrent}, hybrid networks \cite{li2023gdf}, etc., which results in promising performance.

Yan et al. \cite{yan2018three} proposed a three-stage segmentation network that addresses separate segmentation problems for thick and thin vessels in the first two stages and fuses features from these stages in the last stage. Yang et al. \cite{yang2021hybrid} introduced a hybrid fusion network to process thick and thin vessel information separately, with a subsequent fusion network to aggregate these two types of vessel information, achieving high recall on the DRIVE dataset. Wang et al. \cite{wang2020hard} presented a multi-decoder structure to process different types of vascular information by classifying features as 'Hard' or 'Easy', and then processing them separately in subdecoders to adaptively handle different types of features.

In another approach, a dual-path encoder \cite{wang2020csu} was proposed, consisting of two paths that process spatial features and contextual features separately using different sizes of convolutional kernels to capture richer features. Staal et al. \cite{zhang2022bridge} introduced Bridge-Net, which employed U-Net as the baseline network and used two patches at different scales to process vascular features, allowing large-scale patches to provide additional contextual information to small-scale patches. DEF-Net \cite{li2022def} used a dual encoder structure to enrich feature representation, capture detailed features and contextual features through CNN and RCNN structures, respectively, and facilitate feature fusion with multiscale fusion blocks.

While these segmentation methods process thick and thin vessels individually based on their distinct characteristics, the multipath structure introduces an additional computational burden. Moreover, the features of the thin vessels are often lost in successive convolution and pooling operations, resulting in inefficient fusion of thick and thin vessel information and making it challenging to achieve the desired segmentation accuracy.

In response to the aforementioned challenges, we present a novel, lightweight, multipath bidirectional skip connection-based CNN to segment the vascular structures in retinal images. The proposed LMBiS-Net scheme, illustrated in Fig. \ref{fig:Model}, is based on a lightweight architecture. Initially, we introduce the multipath feature extraction block for feature extraction. Additionally, bidirectional skip connections are used to integrate robust feature fusion in the encoder-decoder architecture. Furthermore, the encoder utilizes only two max-pooling layers to minimize spatial information loss. Experiments were carried out on three different datasets focused on segmentation of retinal blood vessels, showcasing the effectiveness of the proposed architecture.

\begin{figure}
    \centering
    \includegraphics[width=\textwidth]{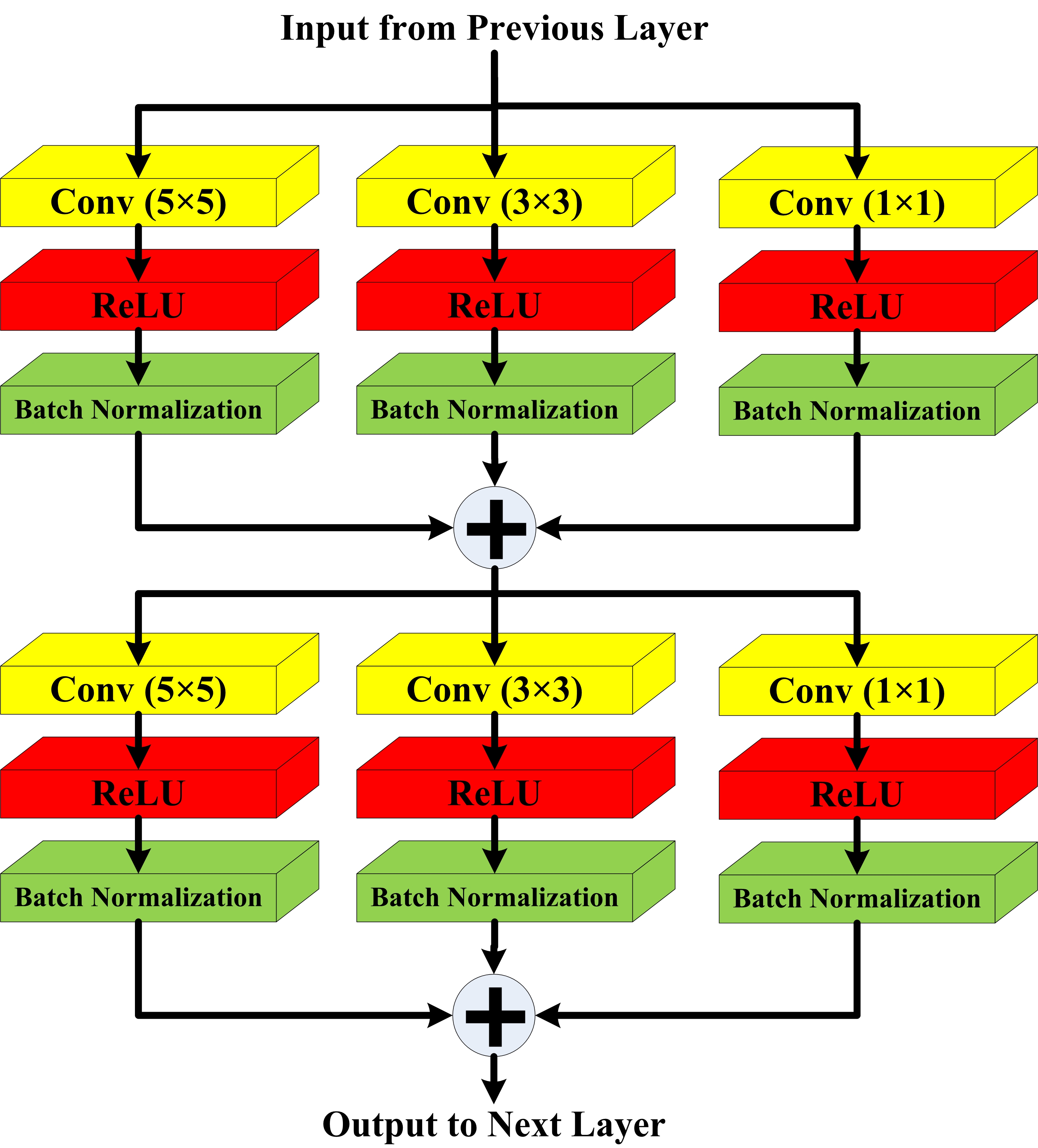}
    \caption{Schematic of the proposed multipath block used for the feature information extraction.}
    \label{fig:Multipathblock}
\end{figure}

\section{Literature Review}
\label{sec:Related Work}
The prevailing body of knowledge encompasses a range of image segmentation methodologies, encompassing techniques such as image thresholding \cite{otsu1979threshold}, k-means clustering \cite{iqbal2022recent}, and methods rooted in Markov random fields \cite{zhou2019review}. However, with the advent of AI and DL, an array of deep learning-based image segmentation strategies has emerged, including encoder-decoder models and FCN \cite{iqbal2022g}.  Deep learning has shown great promise in this area, enabling more accurate and efficient segmentation of biomedical images for improved diagnosis and analysis. Convolutional Neural Networks (CNNs) serve as a pivotal cornerstone in various domains of computer vision, prominently encompassing applications such as image semantic segmentation \cite{kayalibay2017cnn}. The introduction of FCN has been a significant advancement in semantic image segmentation \cite{long2015fully}. Furthermore, an encoder-decoder architecture based on a 13-layer deep neural network has been introduced for semantic segmentation \cite{badrinarayanan2015segnet}. Another notable breakthrough in biomedical image segmentation is the U-Net model, an encoder-decoder architecture that employs skip connections to preserve information and uses data augmentation to synthetically increase the size of dataset  during model training \cite{ronneberger2015u}.

Various adaptations of the U-Net architecture have been extensively explored in the literature for image segmentation tasks, encompassing U-Net++, Attention U-Net, Residual U-Net, Recurrent U-Net and Inception U-Net \cite{siddique2021u}. These U-Net variations have found notable applications within the domain of biomedical imaging, spanning CT, MRI, and X-ray images \cite{du2020medical}. For example, a modified U-Net model with additional residual skip connections was proposed for brain tumor segmentation in \cite{rehman2020bu}. The continuous evolution and customization of U-Net models and their variants have contributed substantially to the precision and efficiency of biomedical image segmentation, unlocking new avenues for medical diagnosis and analysis.

Retinal blood vessel segmentation has been addressed through various methodologies in the existing literature. Ensemble decision tree classifiers were utilized in some studies for image analysis \cite{fraz2012ensemble}, while others employed wavelet-based frameworks to approximate the diameter of the retinal vessel in fundus images \cite{fathi2013automatic}. An innovative deep learning model, Sine-Net, was introduced for retinal vessel segmentation in \cite{atli2021sine}. FCNs have emerged as a common choice for semantic segmentation in this context, with the M3FCN proposal, a variant incorporating a multiscale input block, as described in \cite{jiang2019automatic}. A comprehensive comparative examination of the models of segmentation of DL / ML-based retinal vessels is detailed in \cite{imran2019comparative}. In \cite{mookiah2021review}, the challenges associated with retinal vessel segmentation are discussed alongside a survey of frameworks based on machine learning. Transfer-learning-based FCNs were suggested for retinal vessel segmentation in \cite{jiang2018retinal}. Furthermore, a comprehensive review of multiple CNN-based retinal vessel segmentation networks is presented in \cite{sule2022survey}. These diverse strategies and studies exemplify the ongoing dedication and progress in segmentation of the retinal vessels, aimed at improving medical diagnosis and treatment. The architectural design of Deeplabv3 integrates dilated convolutions and the atrous spatial pyramid pooling module for proficient medical image segmentation \cite{chen2017rethinking}. Subsequently, Chen et al. extended this concept in \cite{chen2018encoder} to introduce Deeplab V3+, an enriched version that encompasses both encoder and decoder components, facilitating the generation of segmented masks. In a different vein, ResDo-UNet, as presented in \cite{liu2023resdo}, augments the U-Net model by incorporating a residual network as the backbone. This modification enhances context capture for retinal vessel segmentation. Similarly, the work of \cite{deshmukh2022retinal} takes advantage of a CNN-based approach for segmentation of retinal vessels in fundus images, coupled with an image quality enhancement strategy to refine the precision of the resulting segmented masks. Moreover, the research conducted in \cite{kar2023retinal} introduces a GAN-based model. This model employs a dual architecture consisting of a generator and discriminator, which work, respectively, to produce segmented masks and perform binary classification. These diverse strategies exemplify the ongoing exploration and advancement of image segmentation techniques, particularly the emphasis on segmentation of retinal vessels. These developments contribute significantly to increasing the efficacy of medical analyses and diagnoses.

In the past, retinal vascular segmentation has been tackled using various approaches, including machine learning, classical image processing techniques, and hybrid methodologies. While some of these methods achieved partial success in accurately segmenting retinal blood vessels, they encountered notable limitations, including low accuracy, sensitivity, specificity, and the requirement for extensive manual parameter tuning. These drawbacks highlight the need for more robust and efficient segmentation techniques to improve the precision and reliability of retinal vascular segmentation in medical image analysis. 

\begin{table*}
  \centering
   \adjustbox {max width=\textwidth}
   {
  \caption{Description of the datasets used for experimentation and evaluation of the proposed LMBiS-Net.}
    \begin{tabular}{lcccccccc}
    \toprule
    \multirow{2}[4]{*}{\textbf{Dataset}} & \multicolumn{3}{c}{\textbf{Number of Images}} & \multicolumn{1}{l}{\textbf{Augmented}} & \textbf{Image } & \textbf{Resized } & \multirow{2}[4]{*}{\textbf{FOV}} & \multirow{2}[4]{*}{\textbf{Image Format}} \\
\cmidrule{2-4}          & \textbf{Training} & \textbf{Testing} & \textbf{Total} & \textbf{Images} & \textbf{Resolution} & \textbf{Resolution} &       &  \\
    \midrule
    DRIVE \cite{DRIVEdata} & 20    & 20    & 40    & 760   & $584 \times 565$ & \multirow{3}[6]{*}{$512\times 512$} & $45^{o}$ & .tif \\
    STARE \cite{STAREDataset} & 16    & 4     & 20    & 608   & $605 \times 700$ &       & $35^{o}$ & .ppm \\
    CHASE \cite{CHASEDataset}& 20    & 8     & 28    & 760   & $990 \times 960$ &       & $30^{o}$ & .jpg \\
    \bottomrule
    \end{tabular}%
    }
  \label{tab:datasets}%
\end{table*}%

\section{Proposed Method}
\label{sec:LMBiS-Net}
\subsection{Network Structure}
The proposed LMBiS-Net follows a structured architecture consisting of three encoder blocks, three decoder blocks, a multipath block and bottleneck layers as shown in \ref{fig:Model}. The details of each component are as follows: Convolution layers and ReLU activation functions are used to extract features from the given input. A batch normalization layer is applied to enhance training stability and accelerate convergence during optimization. After batch normalization, max pooling is employed to reduce the spatial dimensions of the feature maps while retaining essential feature information. The multipath feature extraction block incorporates convolution blocks $1\times 1$, $3\times 3$, and $5\times 5$, each followed by ReLU activation and batch normalization. The output of these blocks is summed to generate a feature map, which represents the output of this encoder stage. The bottleneck layer plays a vital role in training the model while maintaining computational efficiency. It typically acts as a bridge between the encoder and decoder sections.  In the decoder stage, transpose convolution layers are used for up-sampling, effectively increasing the spatial dimensions of the feature maps. Transpose convolution is the reverse operation of standard convolution. At the final decoder stage, a softmax layer is employed for normalization and scale invariance. It produces the model's output in a probabilistic form. Dice pixel classification measures the overlap between the predicted image and the ground truth. It is commonly used in segmentation tasks to assess the similarity between the predicted and actual segmentations. Skip connections are a powerful architectural design in neural networks that enable the training of deep models. The proposed LMBiS-Net integrates bidirectional skip connections connecting its encoding and decoding layers, facilitating the fusion of low-level and high-level feature data using forward skip connections. Furthermore, reverse skip connections are utilized to remap the decoded features to the encoder.

\subsection{Multipath Feature Extraction Block}
A multipath feature extraction block is introduced in the proposed LMBiS-Net to process information through multiple parallel pathways. The key advantage of a multipath feature extraction block is its ability to introduce feature diversity. Each pathway within the block can specialize in capturing distinct features from the input data. This diversity allows the network to learn different types of pattern or detail, resulting in a more comprehensive and varied representation of the given input. By processing the input data through multiple paths with varying levels of complexity, a multipath feature extraction block helps to build a hierarchical representation. This means the network can capture both low-level features (e.g., edges and textures) and high-level features (e.g., complex objects or structures), which is vital for vessel segmentation tasks. Multipath feature extraction blocks encourage diversity among pathways, acting as a form of regularization. This diversity helps prevent overfitting by reducing the risk that the entire network memorizes the training data. It promotes robustness and generalization. The proposed multipath feature extraction block is shown in \ref{fig:Multipathblock} that performs the following operations: The convolutions $1\times 1$, $3\times 3$, and $5\times 5$ are applied to the input data. The ReLU activation function is applied to introduce non-linearity. Batch normalization is performed to stabilize and accelerate training. The outputs from batch normalization in all convolution blocks are aggregated, resulting in the generation of intermediate outputs. This process is repeated in subsequent stages, where the output from the previous stage becomes the input from the next stage. The final output is directed to the bottleneck layer, which is responsible for training the model while maintaining computational efficiency. The input $I_{in}$ is passed through a convolutional block of three different scales, s, and the intermediate output $S_{1}$ is obtained using the equation \ref{Eq:1}.

\begin{equation}
    S_{1}=\sum_{k=1}^{3}\beta \left ( \text{ReLU} \left ( f^{n\times n} \left ( I_{in} \right )\right ) \right )
    \label{Eq:1}
\end{equation}

where $\beta$ is the batch normalization operation, $\text{ReLU}$ is the activation function, $f^{n\times n}$ is the convolution operation of kernel size $n\times n$, and $n=2k-1$.

\begin{equation}
    I_{out}=\sum_{k=1}^{3}\beta \left ( \text{ReLU} \left ( f^{n\times n} \left ( S_{1} \right )\right ) \right )
    \label{Eq:2}
\end{equation}

The final output $I_{out}$ of the multipath feature extraction block is calculated using equation \ref{Eq:2}.

\begin{table*}
  \centering
  \caption{Ablation study performed on DRIVE dataset. $\uparrow$ shows that the higher values are better, whereas $\downarrow$ shows that the lower values are better.}
   \adjustbox {max width=\textwidth}
   { 
    \begin{tabular}{lccccccc}
    \toprule
    \multirow{2}[4]{*}{\textbf{Method}} & \multirow{2}[4]{*}{\textbf{Param (M) $\downarrow$}} & \multicolumn{6}{c}{\textbf{Performance Measures (\%)}} \\
\cmidrule{3-8}          &       & \textbf{$S_{e} \uparrow$} & \textbf{$S_{p} \uparrow$} & \textbf{$A_{cc} \uparrow$} & \textbf{$F_{1}-Score \uparrow$} & \textbf{$AUC \uparrow$} & \textbf{$J_{index} \uparrow$} \\
    \midrule

    Base Line (BL) & 13.00 & 80.54 & 98.06 & 95.33 & 79.53 & 97.12 & 63.94 \\
    Lightweight Base Line(LBL) & 0.095 & 78.10 & 98.18 & 96.42 & 79.21 & 88.14 & 62.44 \\
    LBL + Multipath Feature Extraction Block (MFEB) & 0.150 & 80.35 & 98.28 & 96.70 & 81.00 & 89.31 & 64.35 \\
    LBL+ MFEB + Bidirectional Skip Connections & 0.172 & 83.60 & 98.32 & 96.83 & 81.76 & 89.81 & 65.33 \\
   
    \bottomrule
    \end{tabular}%
    }
  \label{tab:Ablation}%
\end{table*}%

\begin{table}
  \centering
   \caption{A performance comparison between LMBiS-Net and several alternative methods was conducted using the DRIVE \cite{DRIVEdata} dataset. $\uparrow$ shows that the higher values are better.}
    \adjustbox{max width=\textwidth}{%

    \begin{tabular}{lccccc}
    \toprule
    \multirow{2}[4]{*}{\textbf{Method}} & \multicolumn{5}{c}{\textbf{Performance Measures in (\%)}} \\
\cmidrule{2-6}          & \textbf{$S_{e} \uparrow$} & \textbf{$S_{p} \uparrow$} & \textbf{$A_{cc} \uparrow$} & \textbf{$AUC \uparrow$} & \textbf{$F_{1} \uparrow$} \\
    \midrule
    Orlando et al. \cite{orlando2016discriminatively} & 78.97 & 96.84 & 94.54 & 95.06 & - \\
    Att UNet \cite{oktay2018attention} & 79.46 & 97.89 & 95.64 & 97.99 & 82.32 \\
    H-DenseUNet \cite{li2018h} & 79.85 & 98.05 & 95.73 & 98.10 & 82.79 \\
    BTS-DSN \cite{Guo2019} & 78.00 & 98.06 & 95.51 & 97.96 & 82.08 \\
    BCDU-Net \cite{azad2019bi} & 79.84 & 98.03 & 95.75 & 98.11 & 82.49 \\
    Bio-Net \cite{xiang2020bio} & 82.20 & 98.04 & 96.09 & 82.06 & 98.26 \\
    CTF-Net \cite{wang2020ctf} & 78.49 & 98.13 & 95.67 & 97.88 & 82.41 \\
    CSU-Net \cite{wang2020csu} & 80.71 & 97.82 & 95.65 & 98.01 & 82.51 \\
    OCE-Net \cite{OCE-NET} & 80.18 & 98.26 & 95.81 & 98.21 & 83.02 \\
    G-Net Light \cite{iqbal2022g} & 81.92 & 98.29 & 96.86 & -     & 82.02 \\
    LDMRes-Net \cite{iqbal2023ldmres} & 83.58 & 98.32 & 97.02 & 98.51 & 83.09 \\
    \midrule

    \textbf{Proposed LMBiS-Net} & \textbf{83.60} & \textbf{98.83} & \textbf{97.08} & \textbf{98.80} & \textbf{83.43} \\
    \bottomrule
    \end{tabular}%
    }
  \label{tab: DRIVE}%
\end{table}%

\begin{figure*}
	\centering
    \includegraphics[width=\textwidth]{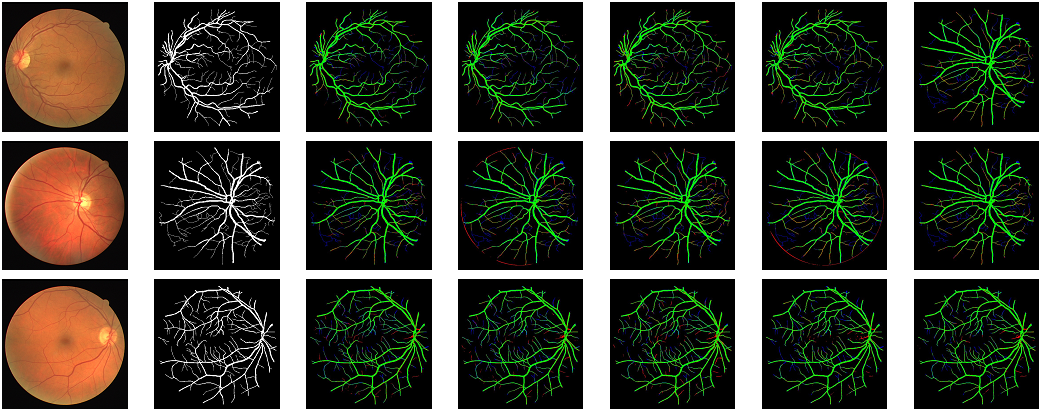}  
	\caption{Segmentation results of LMBiS-Net and comparative methods on representative test images of the DRIVE dataset. From left to right: test image, ground-truth segmentation, and the segmentation results of BCDU-Net, MultiResNet, SegNet, Unet++, and our proposed LMBiS-Net, respectively. True positive pixels are shown in green, false positives in red, and false negatives in blue.}
	\label{visualDRIVE}%
\end{figure*}%

\section{Experimental Setup}\label{experimentalResults}
\subsection{Datasets Description}

The proposed model was evaluated and compared with other models using three publicly available retinal fundus image datasets: DRIVE, STARE, and CHASE\_DB1. The DRIVE dataset comprises 40 fundus images, each with a size of $584\times 565$ pixels. The STARE dataset consists of 20 fundus images with dimensions of $700\times 605$ pixels, while the CHASE\_DB1 dataset includes 28 images with dimensions of $999\times 960$ pixels. Due to the limited number of fundus images in each dataset, data augmentation was applied to increase the dataset size. As a result, the DRIVE dataset was augmented to 240 fundus images, of which 220 were used for training and 20 for testing. The STARE dataset was expanded to 170 fundus images, with 165 images used for training and 5 for testing. Similarly, the CHASE\_DB1 dataset was augmented to 228 fundus images, and the split between training and testing was 220 and 8 images, respectively. For further details, refer to Table ~\ref{tab:datasets}, which provides a summary of the data set used in the study.

\begin{figure*}[h]
	\centering
	\includegraphics[width=\textwidth]{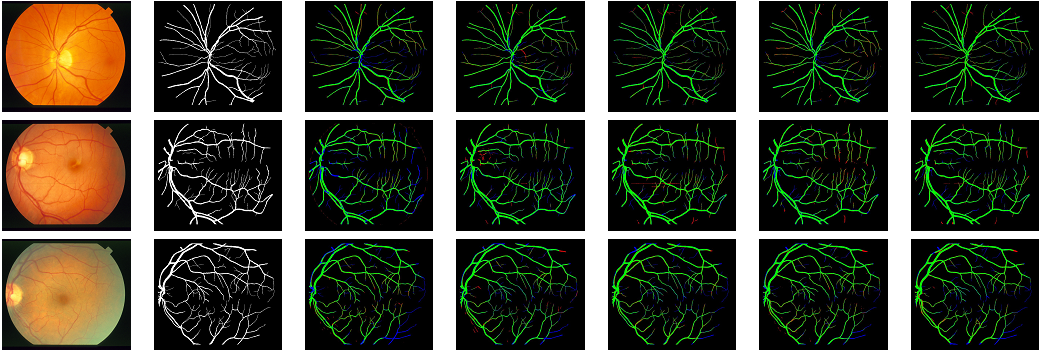}  
	\caption{Segmentation results of LMBiS-Net and comparative methods on representative test images of the STARE dataset. From left to right: test image, ground-truth segmentation, and the segmentation results of BCDU-Net, MultiResNet, SegNet, Unet++, and our proposed LASC-Net, respectively. True positive pixels are shown in green, false positive pixels in red, and false negative pixels in blue.}
 \label{visualSTARE}%
\end{figure*}%

\subsection{Evaluation Criteria}
The evaluation of the segmentation results was performed by comparing them with their corresponding ground-truth images. Each pixel in the output image was classified as correctly segmented foreground pixels ($T_{P}$: true positives) or background pixels ($T_{N}$: true negatives), or as wrongly segmented foreground pixels ($F_{P}$: false positives) or background pixels ($F_{N}$: false negatives).

Sensitivity (Se) is the ratio of ({$T_{P}$}) to the total number of actual positives ({$T_{P}$+ $F_{N}$}). It represents the ability of the method to detect all positive samples correctly. It can be formulated as 

\begin{equation}
S_{e}=\frac{T_{P}}{T_{P}+F_{N}}
\label{eq:sn}
\end{equation}

Specificity (Sp) is the ratio of ({$T_{N}$) to the total number of actual negatives ({$T_{N}$+ $F_{P}$}). It represents the ability of the method to correctly identify all negative samples. It can be formulated as 

\begin{equation}
S_{p}=\frac{T_{N}}{T_{N}+F_{P}}
\label{eq:sp}
\end{equation}

Accuracy (Acc) is the proportion of correctly identified pixels ({$T_{P}$+ $T_{N}$}) to all pixels in the image. It displays the method's total effectiveness. It can be stated as follows. 

\begin{equation}
A_{cc}=\frac{T_{P}+T_{N}}{T_{P}+T_{N}+F_{P}+F_{N}}
\label{eq:acc}
\end{equation}

The $F_{1}$-score, or Dice similarity coefficient (DSC), is another widely used metric to evaluate model performance. It is calculated as the harmonic mean of precision and recall, and can be expressed as:

\begin{equation}
F_{1}-Score =\frac{2 \times T_{P}}{(2\times T_{P})+F_{P}+F_{N}}
\label{eq:F1}
\end{equation} 

The area under the curve (AUC) is a metric used in receiver operating characteristic (ROC) analysis. The ROC curve plots the true positive rate (Se) against the false positive rate ({1+ $S_{p}$})(1-Sp) for different classification thresholds. AUC measures the overall ability of the method to distinguish between positive and negative samples. 

\begin{equation}
AUC=1-\frac{1}{2}\left(\frac{F_{P}}{F_{P}+T_{N}}+\frac{F_{N}}{F_{N}+T_{P}}\right)
\label{eq:AUC}
\end{equation}


\begin{table}
  \centering
    \caption{Performance comparison of LMBiS-Net and a number of alternatives on the STARE \cite{STAREDataset} dataset. $\uparrow$ shows that the higher values are better.}
    \adjustbox{max width=\textwidth}{

    \begin{tabular}{lccccc}
    \hline
    \multirow{2}[4]{*}{\textbf{Method}} & \multicolumn{5}{c}{\textbf{Performance Measures in (\%)}} \\
\cmidrule{2-6}          & \textbf{$S_{e} \uparrow$} & \textbf{$S_{p} \uparrow$} & \textbf{$A_{cc} \uparrow$} & \textbf{$AUC \uparrow$} & \textbf{$F_{1} \uparrow$}  \\
    \hline
    Orlando et al. \cite{orlando2016discriminatively} & 76.80 & 97.38 & 95.19 & 95.70 & - \\
    Att UNet \cite{oktay2018attention} & 77.09 & 98.48 & 96.33 & 97.00 & - \\
    BTS-DSN \cite{Guo2019} & 82.01 & 98.28 & 96.60 & 98.72 & 83.62 \\
    BCDU-Net \cite{azad2019bi} & 78.92 & 98.16 & 96.34 & 98.43 & 82.30 \\
    CC-Net \cite{Feng2020} & 80.67 & 98.16 & 96.32 & 98.33 & 81.36 \\
    OCE-Net \cite{OCE-NET} & 80.12 & 98.65 & 96.72 & 98.76 & 83.41 \\
    Wave-Net \cite{liu2022wave} & 79.02 & 98.36 & 96.41 & -     & 81.40 \\
    G-Net Light \cite{iqbal2022g} & 81.70 & 98.53 & 97.30 & -     & 81.78 \\
    LDMRes-Net \cite{iqbal2023ldmres} & 84.07 & 98.75 & 97.64 & 98.72 & 84.24 \\
    \hline
    \textbf{Proposed LMBiS-Net} & \textbf{84.37} & \textbf{98.77} & \textbf{97.69} & \textbf{98.82} & \textbf{84.44} \\
    \hline
    \end{tabular}%
    }
  \label{tab:STARE}%
\end{table}%

\subsection{Implementation and Training}

This section delineates the training details of the proposed LMBiS-Net on three benchmark datasets (DRIVE, STARE, and CHASE\_DB). All images are standardized to a resolution of $512\times 512$ pixels to ensure consistency in size. Subsequently, to resizing, the images are subjected to enhancement procedures, including contrast adjustments (with factors of [$\times 0.9, \times 1.1$]) and rotation at $10^{o}$. The size of the dataset increases ($38\times$) of the actual size of the datasets after the augmentations. Segmentation models undergo training via dice loss.  Adam optimizer is employed, with a cap of 50 iterations and an introductory learning rate of 0.001. In the absence of performance improvement on the validation set after seven epochs, the learning rate is reduced by a quarter. To stop overfitting, an early stopping strategy is implemented. The LMBiS-Net is implemented through Keras using TensorFlow as the back-end and is trained on a NVIDIA K80 GPU.

\section{Results and Discussions}
\label{sec:Results}
In this section, we will evaluate the overall performance of the proposed LMBiS-Net through a series of experiments. First, we conduct ablation experiments to assess the effectiveness of each proposed network block and validate the contributions of our proposed modifications. Second, we compare the segmentation results of our proposed network with those of some advanced methods. Additionally, we present visualization results to visually demonstrate the superiority of our approach in capturing important structures and details in the input data. Finally, we present the computational complexity analysis of the proposed LMBiS-Net.

\begin{figure*}[h]
	\centering
 
 \includegraphics[width=\textwidth]{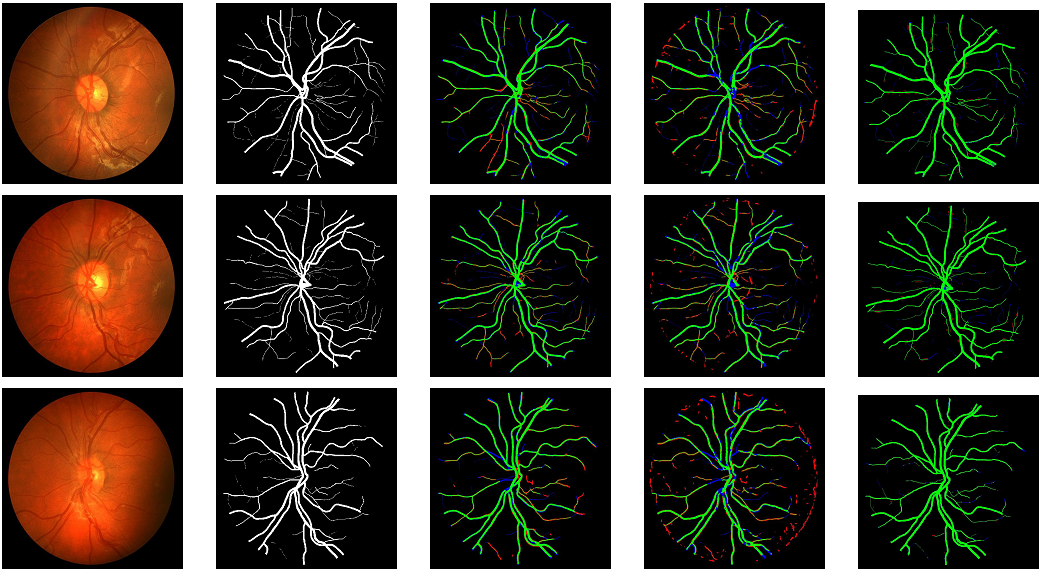}   
	\caption{Segmentation results of LMBiS-Net and comparative methods on representative test images of the CHASE dataset. From left to right: test image, ground-truth segmentation, and the segmentation results of SegNet, VessSeg, and our proposed LMBiS-Net, respectively. True positive pixels are shown in green, false positive pixels in red, and false negative pixels in blue.}\label{visualCHASE}%
\end{figure*}%

\begin{table}
  \centering
    \caption{A performance comparison between LMBiS-Net and several alternative methods was conducted using the CHASE\_DB \cite{CHASEDataset} dataset. $\uparrow$ shows that the higher values are better.}
  \adjustbox{max width=\textwidth}{

    \begin{tabular}{lccccc}
    \toprule
    \multirow{2}[4]{*}{\textbf{Method}} & \multicolumn{5}{c}{\textbf{Performance Measures in (\%)}} \\
\cmidrule{2-6}          & \textbf{$S_{e} \uparrow$} & \textbf{$S_{p} \uparrow$} & \textbf{$A_{cc} \uparrow$} & \textbf{$AUC \uparrow$} & \textbf{$F_{1} \uparrow$} \\
    \midrule
    Orlando et al. \cite{orlando2016discriminatively} & 75.65 & 96.55 & 94.67 & 94.78 & - \\
    Att UNet \cite{oktay2018attention} & 80.10 & 98.04 & 96.42 & 98.40 & 80.12 \\
    BTS-DSN \cite{Guo2019} & 78.88 & 98.01 & 96.27 & 98.40 & 79.83 \\
    BCDU-Net \cite{azad2019bi} & 77.35 & 98.01 & 96.18 & 98.39 & 79.32 \\
    OCE-Net \cite{OCE-NET} & 81.38 & 98.24 & 96.78 & 98.72 & 81.96 \\
    G-Net Light \cite{iqbal2022g} & 82.10 & 98.38 & 97.26 & -     & 80.48 \\
    LDMRes-Net \cite{iqbal2023ldmres} & 85.95 & 98.88 & 97.55 & 98.61 & 81.94 \\
    
    \midrule
    
    \textbf{Proposed LMBiS-Net} & \textbf{86.05} & \textbf{98.96} & \textbf{97.75} & \textbf{98.71} & \textbf{83.54} \\
    \bottomrule
    \end{tabular}%
    }
  \label{tab:CHASE}%
\end{table}%

\subsection{Ablation Study on DRIVE dataset}

An ablation study of the proposed LMBiS-Net is performed on the DRIVE\cite{DRIVEdata} dataset. Experiments are conducted in a progressive manner. We have started with implementing a single-path encoder-decoder Unet \cite{ronneberger2015u} network. After that, we reduced the number of parameters to make a lightweight version of Unet by reducing the number of filters at each convolution layer and the encoder depth by minimizing the pooling operations. This reduction of parameters leads to a performance decrease, which is then enhanced by the use of a multipath feature extraction block (MFEB). After that, bidirectional skip connections are employed between the encoder and decoder to improve the information flow and refinement of the extracted feature in the reconstruction phase of the proposed LMBiS-Net. Table \ref{tab:Ablation} shows the quantitative results of the ablation study carried out and it is evident that performance is significantly improved while having very few learnable parameters.

\begin{table*}
  \centering
  \caption{Cross validation of the proposed LMBiS-Net across the DRIVE, STARE, and CHASE\_DB datasets. $\uparrow$ shows that the higher values are better.}
  \adjustbox{max width=\textwidth}{
  
       \begin{tabular}{lllccccc}
    \toprule
    \multirow{2}[4]{*}{\textbf{Training Dataset}} & \multicolumn{1}{c}{\multirow{2}[4]{*}{\textbf{Testing Dataset}}} & \multirow{2}[4]{*}{\textbf{Method}} & \multicolumn{5}{c}{\textbf{Performance Measures in (\%)}} \\
\cmidrule{4-8}          &       &       & \textbf{$S_{e} \uparrow$} & \textbf{$S_{p} \uparrow$} & \textbf{$A_{cc} \uparrow$} & \textbf{$AUC \uparrow$} & \textbf{$F_{1} \uparrow$}  \\
    \midrule
    \multirow{8}[8]{*}{\textbf{DRIVE}} & \multirow{4}[4]{*}{\textbf{CHASE\_DB}} & BCDU-Net \cite{azad2019bi} & 72.17 & 98.20 & 94.86 & 93.27 & 73.67 \\
          &       & G-Net Light \cite{iqbal2022g} & 72.92 & 98.15 & 95.02 & 97.09 & 73.13 \\
          &       & Att UNet \cite{oktay2018attention} & 75.36 & 98.62 & 94.65 & 97.32 & 75.22 \\
\cmidrule{3-8}          &       & \textbf{LMBiS-Net} & \textbf{79.02} & \textbf{98.70} & \textbf{94.95} & \textbf{97.46} & \textbf{78.56} \\
\cmidrule{2-8}          & \multicolumn{1}{l}{\multirow{4}[4]{*}{\textbf{STARE}}} & BCDU-Net \cite{azad2019bi} & 74.99 & 97.98 & 95.63 & 96.21 & 75.06 \\
          &       & G-Net Light \cite{iqbal2022g} & 77.22 & 97.15 & 95.48 & 94.86 & 77.98 \\
          &       & Att UNet \cite{oktay2018attention} & 75.06 & 98.28 & 95.53 & 96.23 & 76.89 \\
\cmidrule{3-8}          &       & \textbf{LMBiS-Net} & \textbf{79.56} & \textbf{98.31} & \textbf{95.89} & \textbf{97.06} & \textbf{80.22} \\
    \midrule
    \multirow{8}[8]{*}{\textbf{STARE}} & \multirow{4}[4]{*}{\textbf{DRIVE}} & BCDU-Net \cite{azad2019bi} & 75.34 & 97.23 & 95.07 & 93.27 & 76.87 \\
          &       & G-Net Light \cite{iqbal2022g} & 76.77 & 97.15 & 95.21 & 95.32 & 77.02 \\
          &       & Att UNet \cite{oktay2018attention} & 75.36 & 98.03 & 95.68 & 96.07 & 76.23 \\
\cmidrule{3-8}          &       & \textbf{LMBiS-Net} & \textbf{77.38} & \textbf{98.70} & \textbf{95.79} & \textbf{97.46} & \textbf{78.01} \\
\cmidrule{2-8}          & \multicolumn{1}{l}{\multirow{4}[4]{*}{\textbf{CHASE\_DB}}} & BCDU-Net \cite{azad2019bi} & 77.29 & 97.65 & 94.88 & 95.09 & 78.25 \\
          &       & G-Net Light \cite{iqbal2022g} & 76.22 & 96.99 & 94.76 & 95.24 & 77.04 \\
          &       & Att UNet \cite{oktay2018attention} & 77.01 & 97.29 & 94.03 & 96.11 & 77.54 \\
\cmidrule{3-8}          &       & \textbf{LMBiS-Net} & \textbf{78.99} & \textbf{98.02} & \textbf{95.66} & \textbf{96.92} & \textbf{78.52} \\
    \midrule
    \multirow{8}[8]{*}{\textbf{CHASE\_DB}} & \multirow{4}[4]{*}{\textbf{DRIVE}} & BCDU-Net \cite{azad2019bi} & 72.17 & 98.20 & 94.86 & 93.27 & 73.32 \\
          &       & G-Net Light \cite{iqbal2022g} & 72.92 & 98.15 & 95.02 & 97.09 & 74.59 \\
          &       & Att UNet \cite{oktay2018attention} & 75.36 & 98.62 & 94.65 & 97.32 & 75.99 \\
\cmidrule{3-8}          &       & \textbf{LMBiS-Net} & \textbf{78.82} & \textbf{98.44} & \textbf{95.55} & \textbf{97.34} & \textbf{79.07} \\
\cmidrule{2-8}          & \multicolumn{1}{l}{\multirow{4}[4]{*}{\textbf{STARE}}} & BCDU-Net \cite{azad2019bi} & 74.99 & 97.98 & 95.63 & 96.21 & 75.01 \\
          &       & G-Net Light \cite{iqbal2022g} & 71.88 & 98.16 & 95.48 & 94.86 & 72.03 \\
          &       & Att UNet \cite{oktay2018attention} & 75.06 & 98.28 & 95.53 & 96.23 & 76.11 \\
\cmidrule{3-8}          &       & \textbf{LMBiS-Net} & \textbf{76.19} & \textbf{98.45} & \textbf{95.71} & \textbf{96.82} & \textbf{76.22} \\
    \bottomrule
    \end{tabular}%
    }
  \label{tab:Cross_Data_Comp}%
\end{table*}%

\subsection{LMBiS-Net Performance Evaluation and Comparison with State-of-the-art Methods}

In this section, we demonstrate the performance comparison of the proposed LMBiS-Net by evaluating its segmentation performance on multiple datasets of retinal vessel segmentation. The datasets utilized in this comparison are DRIVE \cite{DRIVEdata}, STARE \cite{STAREDataset}, and CHASE\_DB1 \cite{CHASEDataset}. For a complete understanding of the performance, we also refer to previous studies that have compared numerous supervised methodologies. In particular, we evaluate the performance of LMBiS-Net against U-Net \cite{Ronneberger2015} and SegNet \cite{Badrinarayanan2017}. These two deep learning architectures are popular and frequently serve as benchmark models within the image segmentation community. 

Tables \ref{tab: DRIVE}, \ref{tab:STARE}, and \ref{tab:CHASE} provide the quantitative evaluation of our proposed LMBiS-Net alongside several other methodologies. As the tables illustrate, LMBiS-Net exceeds all other methods in all metrics, including sensitivity, specificity, accuracy, area under the curve (AUC), and $F1$ score. Specifically, in the DRIVE dataset \cite{DRIVEdata}, LMBiS-Net achieved performance metrics of 83.60\%, 98.83\%, 97.08\%, 98.80\%, and 83.43\% for sensitivity, specificity, precision, AUC and $F1$ score, respectively. Similarly, on the STARE dataset \cite{STAREDataset}, LMBiS-Net achieved values of 84.37\%, 98.77\%, 97.69\%, 98.82\%, and 84.44\% for sensitivity, specificity, precision, AUC, and $F1$ score, respectively. On the CHASE\_DB1 dataset, LMBiS-Net attained scores of 86.05\%, 98.96\%, 97.75\%, 98.71\% and 82.04\% for sensitivity, specificity, precision, AUC, and $F1$ score, respectively. These findings underscore the superior performance of LMBiS-Net compared to other state-of-the-art networks. It is worth noting that among the alternative methods, there is no apparent trend or consistent superior performance in terms of specificity, AUC, or sensitivity.

In addition to the quantitative analysis, we present a qualitative comparison of the results obtained by LMBiS-Net and other methods on the DRIVE, STARE, and CHASE datasets. The results obtained on the DRIVE dataset (Fig.~\ref{visualDRIVE}) demonstrate that LMBiS-Net significantly reduces false positives in small vessels compared to current methods. For example, U-Net variants struggle to accurately delineate vessel boundaries, resulting in a higher number of false positives, while SegNet \cite{Badrinarayanan2017} tends to generate false tiny vessels in most images, and BCDU-Net \cite{azad2019bi} appears to overlook crucial information about vessel structures, leading to suboptimal segmentation performance. In contrast, LMBiS-Net effectively captures this information while minimizing the generation of false vessel information, resulting in more accurate segmentations.

Alternative methods tend to produce more false positives when applied to the STARE dataset (Fig.~\ref{visualSTARE}), particularly around the retinal boundaries, optic nerves, and small vessels. This may be attributed to the challenges posed by the complex retinal structures and image artifacts present in this dataset. The proposed LMBiS-Net method proves to be more robust against these artifacts, preserving the fine details of the vessel structures while maintaining a low rate of false positives. These findings indicate that LMBiS-Net is capable of capturing the subtle characteristics of retinal vessels in challenging scenarios, showcasing its efficacy in this domain.

Similar observations are made when applying LMBiS-Net to the CHASE dataset (Fig.~\ref{visualCHASE}). Despite the presence of various challenges, such as image quality variations and vessel abnormalities, LMBiS-Net consistently achieves accurate vessel segmentation. The proposed method effectively suppresses false positives while preserving the true vessel structures, even in regions with low contrast or overlapping vessel patterns. These results further establish the robustness and efficacy of LMBiS-Net in a variety of datasets and image conditions.

In general, the comprehensive evaluation demonstrates the superiority of our proposed LMBiS-Net method in terms of accuracy and robustness compared to the alternative methods examined. Quantitative analysis reveals consistent high performance, while visual comparisons highlight the ability of LMBiS-Net to accurately capture retinal vessel structures with minimal false positives. These results provide solid evidence for the efficacy of our proposed method for retinal vessel segmentation tasks and its potential to help diagnose and monitor retinal diseases.

To assess the generalizability of our proposed LMBiS-Net, we performed cross-validation experiments on the DRIVE, STARE, and CHASE\_DB datasets and compared the results with other state-of-the-art methods. Performance metrics, including sensitivity ($S_{e}$), specificity ($S{p}$), area under the curve ($AUC$), $F_{1}$, and accuracy ($A_{cc}$), were evaluated. As depicted in Table \ref{tab:Cross_Data_Comp}, the proposed LMBiS-Net demonstrates the best performance in terms of all performance indicators compared to other methods. These findings indicate the superior generalizability of the proposed LMBiS-Net with very few learnable parameters.

\begin{table}[h]
  \centering
  \caption{Computational requirements comparisons of the LMBiS-Net with state-of-the-art methods. $\downarrow$ shows that the lower values are better.}
    \begin{tabular}{lcc}
    \toprule
    Method & \multicolumn{1}{c}{Param (M) $\downarrow$} & \multicolumn{1}{c}{Size (MB) $\downarrow$} \\
    \midrule
    MobileNet-V3-small\cite{cheng2019msnet} & \multicolumn{1}{c}{2.50} & \multicolumn{1}{c}{11} \\
    ERFNet\cite{romera2017erfnet} & \multicolumn{1}{c}{2.06} & \multicolumn{1}{c}{8} \\
    MultiRes UNet\cite{ibtehaz2020multiresunet} & \multicolumn{1}{c}{7.20} & \multicolumn{1}{c}{-} \\
    VessNet\cite{arsalan2019aiding} & \multicolumn{1}{c}{9.30} & \multicolumn{1}{c}{36.6} \\
    PLVS-Net\cite{arsalan2022prompt} & \multicolumn{1}{c}{1.00} & \multicolumn{1}{c}{3.60} \\
    M2U-Net\cite{laibacher2018m2u} & \multicolumn{1}{c}{0.55} & \multicolumn{1}{c}{2.20} \\
    SegR-Net\cite{ryu2023segr} & \multicolumn{1}{c}{0.64} & \multicolumn{1}{c}{-} \\
    G-Net Light\cite{iqbal2022g} & \multicolumn{1}{c}{0.39} & \multicolumn{1}{c}{1.52} \\
    \midrule
    \textbf{Proposed LMBiS-Net} &    \textbf{0.172 }   & \textbf{0.56}  \\
    \bottomrule
    \end{tabular}%
  \label{tab:Parameters}%
\end{table}%

\subsection{Computational Complexity Analysis of the LMBiS-Net}

Table \ref{tab:Parameters} presents a comprehensive comparative analysis based on the model parameters. In deep learning-based medical image analysis, achieving high performance with fewer parameters is a highly desirable aspect, as it leads to faster training and inference times, as well as reduced computational resource requirements. The model with the least number of parameters is highlighted by bold values in Table \ref{tab:Parameters}. The analysis reveals that LMBiS-Net exhibits superior performance compared to other models but has fewer parameters (only 0.172 M), which indicates its efficiency and effectiveness in retinal vessel segmentation. This finding has significant implications for the advancement of automated retinal disease diagnosis tools and the delivery of high-quality healthcare services. With the adoption of LMBiS-Net, clinicians can quickly and accurately analyze retinal images, allowing the earlier detection and treatment of retinal diseases. The efficiency of the proposed model in achieving excellent segmentation results with reduced parameters opens up new possibilities for optimizing medical image analysis and improving patient care in the field of ophthalmology.

\section{Conclusions}
\label{sec:Conclusions}

In this study, we have presented a retinal image segmentation model named the lightweight multipath bidirectional skip connection-based network, which comprises a mere 0.172 million learnable parameters. The proposed LMBiS-Net is designed with a multipath feature extraction block, which significantly improves the network’s computational efficiency. The proposed network uses an optimum number of filters to prevent feature overlap. This significantly reduces training time. The proposed network incorporates bidirectional skip connections for better information flow between the encoder and the decoder for the refinement of encoder features. The experimental findings of the LMBiS-Net on various retinal image datasets of retinal blood confirm the robustness and versatility of the proposed network. LMBiS-Net outperforms all other retinal blood vessel segmentation methods with a merely low number of learnable parameters which is beneficial for early detection and treatment of retinal diseases.



%
%
%

\end{document}